\documentclass[
preprint, showpacs, showkeys
 amsmath,amssymb,
 aps,
prc,
 longbibliography,
 lengthcheck,%
]{revtex4-1}

\usepackage{graphicx}% Include figure files
\usepackage{dcolumn}% Align table columns on decimal point
\usepackage{bm}% bold math
\usepackage{hyperref}% add hypertext capabilities
\begin{document}

%\preprint{APS/123-QED}

\title{Remarks on the non-equilibrium effects and
collision dynamics in heavy-ion collisions}

\author{Yogesh K. Vermani}
\author{Mandeep Kaur}%
\affiliation{Department of Physics, Panjab University, Chandigarh
-160 014, India}

% \collaboration{MUSO Collaboration}%\noaffiliation

%\affiliation{
%\affiliation{
%\affiliation{%

\date{\today}% It is always \today, today,

\begin{abstract}
We study the beam energy dependence of equilibration process and
space-time characteristics of participant and spectator matter.
For this, we simulated the semi-central collisions of $^{40}Ca+
^{40}Ca$ at incident energies of 400, 600 and 1000 AMeV within the
\emph{quantum molecular dynamics} (QMD) approach. Our numerical
calculations based on the molecular dynamics approach show that
incident energy of the projectile influences the reaction
observables drastically. The effect is more visible for transverse
expansion of the nuclear matter and transparency behavior. The
degree of thermalization of participant matter, however, remains
independent of the incident energy. The characteristics of the
trajectories followed by the nucleons suffering maximal and
minimal binary collisions are also analyzed.

\end{abstract}


\maketitle

%\tableofcontents

\section{\label{intro}Introduction}

The simulation of heavy-ion collisions at intermediate energies
has always played a pivotal role in understanding the reaction
mechanism \cite{verm, epl, jpg, prc, sood, bimod, bimod1}, nature
of nucleon-nucleon (\emph{n-n}) interactions, as well as
thermalization achieved by the nuclear system
\cite{dhawanb,mor,fuchs}. Similar efforts are also made at lower
tail of the incident energy where fusion and cluster-decay
processes are dominant \cite{dutt81a,epjrk,dutt81,dutt}. The
choice of various transport codes, beyond doubt, provides a unique
platform to study the properties of hot \& dense nuclear matter
formed during these collisions. In these transport models, one
starts from the well separated projectile and target nuclei and
then evolves towards non-equilibrium stage due to the overlapping
of two Fermi spheres with large relative momentum. The dynamics of
heavy-ion (HI) collision can be interpreted in terms of useful
reaction observables such as collective flow
\cite{leh,mages,wolf,dhawana}, multifragment-emission \cite{verm,
wolf, puri}, temperature and entropy production \cite{wolf},
momentum anisotropy, subthreshold particle production as well as
elliptical flow \cite{dhawanb, wolf, puri, huang, batko, rkp94}.

The different studies have shown that for a given colliding
geometry, the final particle abundances are sensitive towards the
incident energy of the projectile \cite{braun, andro}. These
theoretical predictions, however, were largely based on the
assumption of thermal equilibrium of the nuclear system. It may be
emphasized that at SIS energies, particles are produced in highly
non-equilibrium situation much different from the saturation
characteristics of the normal nuclear matter \cite{sehn}.
Recently, the BUU simulation of $^{197}Au+^{197}Au$ collisions at
1 AGeV reported that spectator matter evolution is strongly
influenced by the compression and expansion of the participant
matter \cite{shi}. The transverse expansion of the spectator
matter is found to be sensitive towards the nuclear
incompressibility of the participant zone \cite{shi}. At
relativistic bombarding energies, the spectator matter
fragmentation \cite{sood, bimod} and its universality
characteristics \cite{verm,poch,schut,gait} have been extensively
studied in recent literature. As far as particpant matter physics
is concerned, extensive study has been made in the past to infer
the properties of hot and dense participant zone \cite{sood,
dhawanb,doss,hadad,liu,yugs}. Unfortunately, excited nuclear
matter under the extreme conditions of temperature and density
exists for a very short duration. This makes it difficult to
extract information about the nuclear forces and reaction dynamics
in experiments. For instance, nuclear matter at nuclear density
$\rho>$1.5$\rho_{\circ}$ lasts for the time span less than 20 fm/c
\cite{rkp94,chomaz1}. It is interesting to study the behavior of
participant matter and its response towards the evolution of the
spectator matter during the collision process. This study also
bears relevance due to its relation with the nuclear equation of
state (EoS) and flow systematics.

During the collision process, different kinds of interactions are
at work which are important in their own \cite{jian,chomaz}. One
is also interested to know behavior of nucleons in the presence of
other nucleons in the surroundings. From the trajectories
traversed by the nucleons in phase space, one can gain important
information on the nature of the hadronic forces as well. In the
present work, we aim to analyze the reaction dynamics in the
semicentral collisions of $^{40}Ca+ ^{40}Ca$ at incident energies
of 400, 600 and 1000 AMeV respectively. We shall also examine the
behavior of nucleons facing maximal and minimal number of
collisions via stopping and trajectories followed in the phase
space during the collision process.

For the simulation of heavy-ion reactions, we shall utilize the
\emph{quantum molecular dynamics} (QMD) model \cite{verm, sood,
bimod,puri, aich, lehm,skm} as primary transport theory which is
described in \ref{model}. Section \ref{result} deals with the
calculations and illustrative results, which are summarized in
section \ref{summary}.

\section{\label{model} The Model}

The quantum molecular dynamics is an N-body theory that simulates
the heavy-ion reactions at intermediate energies on event by event
basis. This is based on a molecular dynamics picture where
nucleons interact via two and three-body interactions. The
explicit two and three-body interactions preserve the fluctuations
and correlations which are important for \emph{N}-body phenomenon.
In this approach, each single nucleon of the two colliding nuclei
is described by a Gaussian wave packet in momentum and coordinate
space. The centroids of these Gaussian wave packets propagate in
coordinate (${\cal R}_3$) and momentum (${\cal P}_3$) spaces in
accordance with the classical equations of motion:
\begin{equation}
\dot{{\bf p}_i}=-\frac{\partial U_{i}}{\partial {\bf r}_i}, ~
\dot{{\bf r}_i}=\frac{{\bf p}_i}{\sqrt{{{\bf p}_i}^{2}+
{m_{i}}^{2}}} + \frac{\partial U_{i}}{\partial {\bf p}_i}.
\label{euler}
\end{equation}
In the quantum molecular dynamics approach, nucleons move along
classical trajectories and undergo stochastic scattering.
Stochastic means that scattering amplitude doesn't relate the
scattering angle with the impact parameter in a unique way like
the collision of classical billiard balls. Collisions are Pauli
blocked if scattered nucleons enter the phase space region which
is already occupied. The local Skyrme interaction used here,
consists of two and three-body \emph{n-n} interactions:
\begin{equation}
V_{ij}^{loc}= t_{1}\delta({\bf r}_i-{\bf r}_j) + t_{2}\delta({\bf
r}_i-{\bf r}_j)\delta({\bf r}_i-{\bf r}_k). \label{pot}
\end{equation}
This interaction (\ref{pot}) is supplemented with a long-range
Yukawa and an effective charge Coulomb interaction parts
\cite{aich}:
\begin{eqnarray}
V_{ij}^{Yuk}&=&t_3 \frac{exp \{ -| {\bf r}_i-{\bf
r}_j|\}/\mu}{|{\bf r}_i-{\bf r}_j|/\mu} \label{yuk}\\
V_{ij}^{Coul}&=& \frac {{Z_i}\cdot{Z_j}~e^2}{|{\bf r}_i -{\bf
r}_j|} \label{col}
\end{eqnarray}
$Z_{i}$, $Z_{j}$ are the effective charge of baryons {\it i} and
{\it j}. In QMD model, one neglects the isospin dependence of the
interaction. All nucleons in a nucleus are assigned the effective
charge $Z=\frac{Z_{T}+Z_{P}}{A_{T}+A_{P}}$ \cite{aich}. The
long-range Yukawa force is necessary to improve the surface
properties of the interaction. The parameters $\mu, t_{1}, t_{2},
t_{3}$ in Eqs (\ref{pot}), (\ref{yuk}) and (\ref{col}) are
adjusted and fitted so as to achieve the correct binding energy
and mean square root values of the radius of the nucleus
\cite{aich}. The Skyrme-type potential can be generalized to
density dependent form as \cite{aich, skm}:
\begin{equation}
U^{Skm} =\alpha (\frac{\rho }{\rho _{o} })+\beta \left(\frac{\rho
}{\rho _{o} } \right)^{\gamma}, \label{sky}
\end{equation}
with $\alpha$ and $\beta$ as free parameters. These are fixed by
the ground state properties of normal nuclear matter. The third
parameter $\gamma$ allows to fix the incompressibility of the
nuclear matter. The two parameterizations are commonly used namely
`soft' and `hard' corresponding to different incompressibilities
of the nuclear matter. We shall use here a `soft' equation of
state with incompressibility ${\cal K}$=200 MeV.

\section{\label{result} Numerical calculations and discussion}

First of all, we study the reaction observables for the
semicentral collisions of $^{40}Ca+ ^{40}Ca$  at incident energies
of 400, 600 and 1000 AMeV. Figure 1 displays the model
calculations of the average nucleon density $\rho^{avg}$,
collision rate $dN_{coll}/dt$, directed transverse momentum
$\langle {p}^{dir}_{x} \rangle$ and momentum anisotropy ratio
$\langle R_{iso} \rangle$ of the participant matter obtained for
the reaction $^{40}Ca+ ^{40}Ca$ at a `reduced' impact parameter
$b/b_{max}$=0.3. The nuclear matter density is calculated as:
\begin{equation}
{\rho}^{avg} =\left \langle \frac{1}{N}
\sum_{i=1}^{N}\sum_{j>i}^{N}\frac{1}{(2\pi L)^{3/2}} e^{-({\bf
r}_{i}(t)-{\bf r}_{j}(t))^{2}/2L} \right \rangle,
\end{equation}
In our approach, the average nuclear matter density $\rho^{avg}$
is calculated in a volume cell of 2 $fm$ with center located at
the point where two nuclei initially touch each other in c.m.
frame. As expected, one obtains maximum central density at the
highest considered energy of 1000 AMeV. A similar behavior can be
seen for the allowed collision rate $dN_{coll}/dt$. The directed
transverse momentum ${p}^{dir}_{x}$ is another important quantity
describing the transverse expansion of the nuclear matter. It is
defined as:
\begin {equation}
\langle{p_{x}^{dir}}\rangle = \frac{1} {N}\sum_{i=1}^{N}~sign(
y_{i})\cdot{p_x}(i),
\end {equation}
with $p_{x}(i)$ as $x$-component of momentum. $y_{i}$ is the
rapidity of $i^{th}$ particle evaluated as:
\begin {equation}
y_{i}= \frac{1}{2} \ell n \frac{{E}(i)+ {p}_{z}(i)
c}{{E}(i)-{p}_{z}(i) c}.
\end {equation}
Here $E(i)$ and ${p}_{z}(i)$ are the total energy and the
longitudinal momentum of the $i^{th}$ particle, respectively.
\begin{figure}[t!]
\centering \vskip -1.0cm
\includegraphics [scale=0.5] {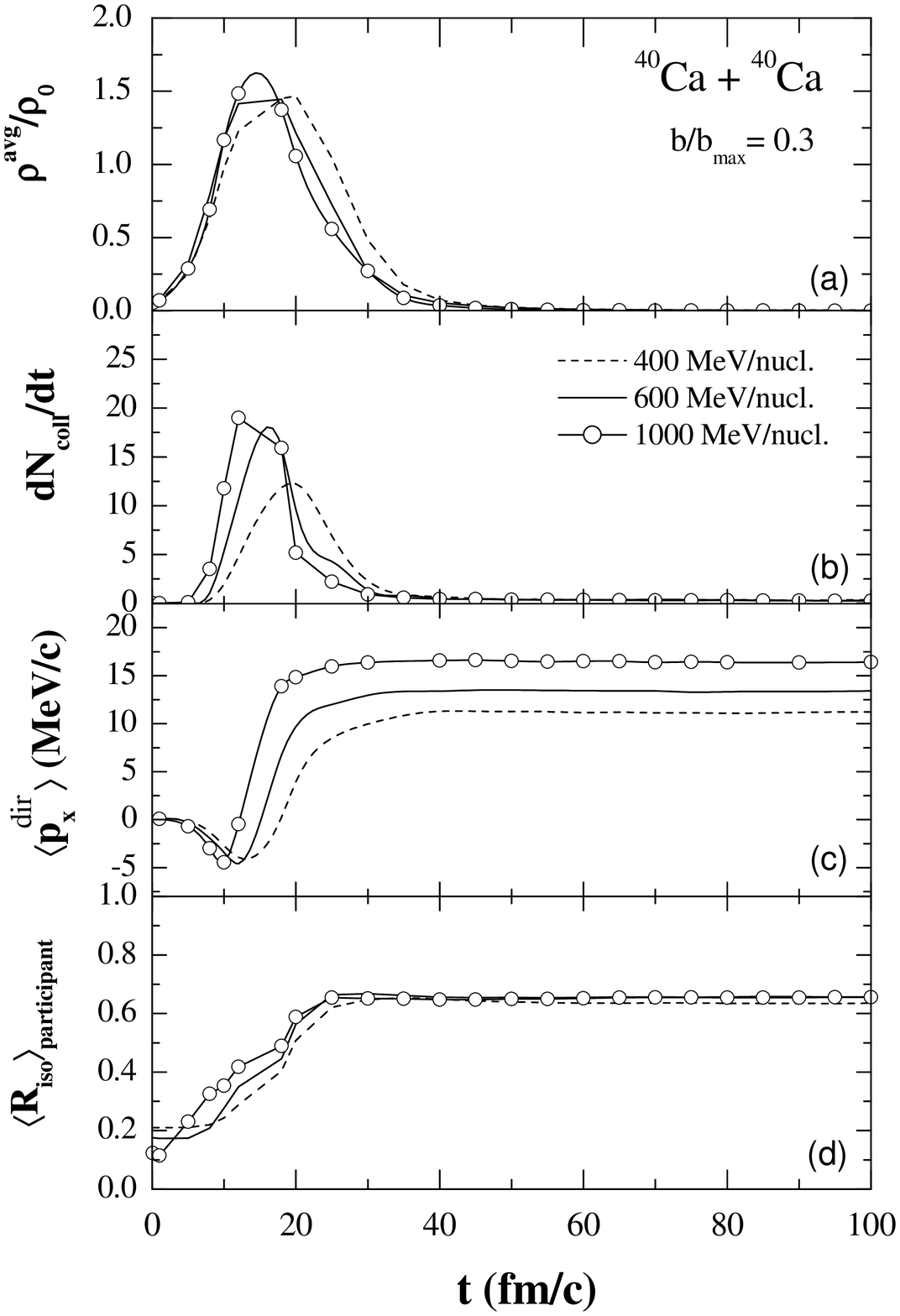} % Here is how to import EPS art
\vskip -0.55cm \caption{The time evolution of $^{40}Ca+ ^{40}Ca$
reaction at incident energies of 400, 600 and 1000 AMev and at
`reduced' impact parameter $b/b_{max}$=0.3. Results are shown here
for: (a) mean central density $\rho^{avg}/\rho_{\circ}$; (b)
average collision rate $dN_{coll}/dt$; (c) directed transverse
momentum $\langle {p}^{dir}_{x} \rangle$; and (d) anisotropy ratio
$\langle R_{iso} \rangle$ of the participant zone.} \label{time}
%\vskip -0.4 cm
\end{figure}
\begin{figure*}[!t]
%\centering
\vskip -2.50cm \includegraphics* [scale=0.62, trim=25 10 0 0, angle=-90] {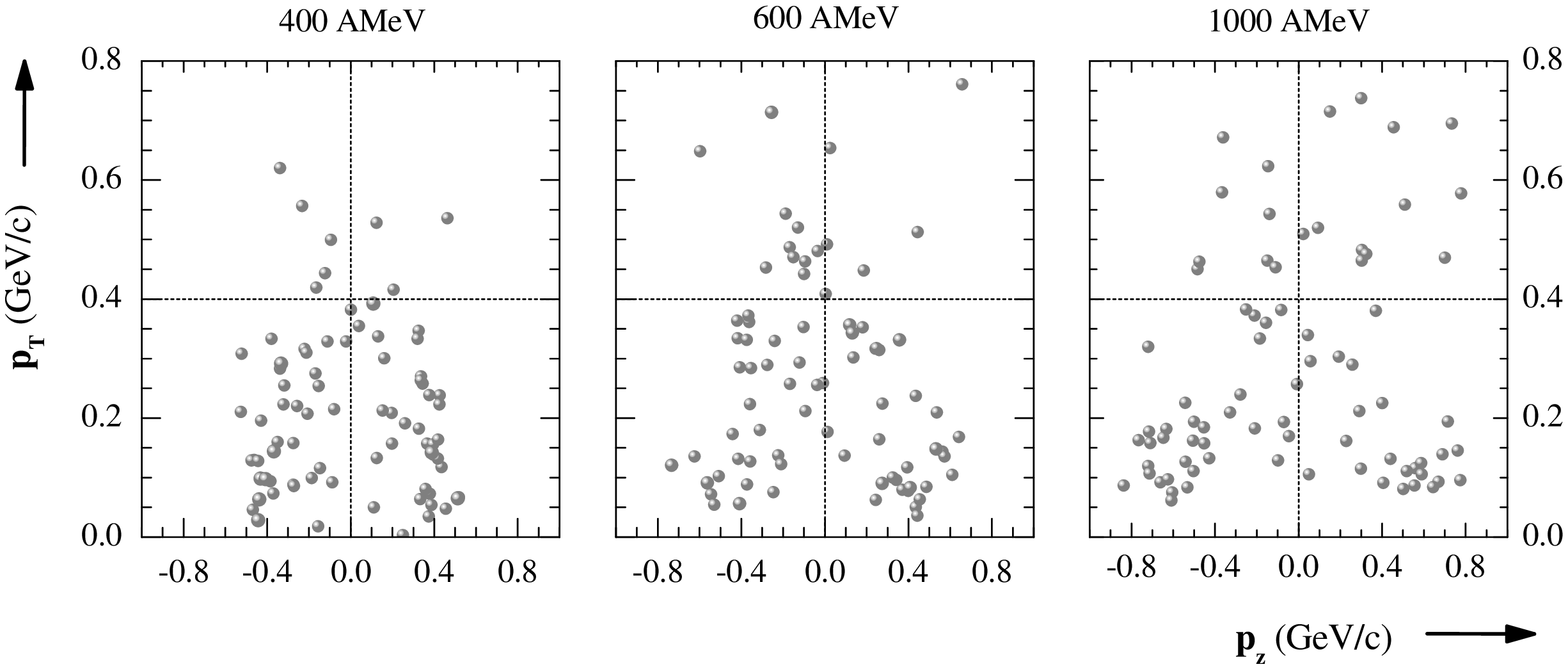} % Here is how to import EPS art
\vskip -2.7cm  \caption {The plots of transverse momentum $p_{T}$
vs longitudinal momentum $p_{z}$ of the nucleons obtained in a
single event of Ca+Ca collision at 400 (left), 600 (middle), and
1000 (right) AMeV and with $b/b_{max}$=0.3. Momentum distribution
of nucleons has been calculated at 100 fm/c.} \label{pt}
%\vskip -0.4 cm
\end{figure*}
Interestingly at the start of the reaction (for $t < 20 fm/c$),
$\langle {p}^{dir}_{x} \rangle$ turns negative (See figure
\ref{time}(c)). This happens due to attractive interaction between
two nuclei in the beginning. After the two nuclei collide, strong
repulsion takes place in the overlap region. As a result,
transverse flow turns positive and saturates finally. The time
evolution of the transverse momentum transfer depicts that it
starts to have finite value when nuclear matter is still
interacting and density is quite high. Beyond 40 fm/c, the
transfer of  momentum in transverse direction is not much altered,
indicating the expansion of the nuclear matter in transverse
direction has ceased. One clearly obtains more transverse
expansion at 1000 AMeV as compared to lower incident energies. It
means that nucleons with larger relative momenta are directed into
the transverse direction with higher momenta. It is worth
interesting to compare the degree of equilibrium attained by the
participant zone at these bombarding energies. We display in the
bottom panel of figure \ref{time}, the momentum anisotropy ratio
$\langle R_{iso} \rangle$ for the participant zone, calculated as
\cite{dhawanb,rkp94}:
\begin{equation}
\langle R_{iso} \rangle = \frac{ \sqrt{ \langle p_{x}^{2} \rangle}
+ \sqrt{\langle p_{y}^{2} \rangle}}{2\sqrt{\langle
p_{z}^{2}\rangle }}. \label{ar}
\end{equation}

\noindent To select momentum anisotropy due to participant zone,
we impose the condition $(-0.5 y_{beam} \leq y \leq 0.5
y_{beam})_{c.m.}$ on the event rapidity. The application of this
cut allows to exclude spectator like components. The full
equilibrium corresponds to $\langle R_{iso} \rangle$ value close
to 1. One can see that equilibration of the participant matter
gets saturated shortly after decompression. This is also the time
when spectator components start departing from the hot interaction
zone and we have the uniform directed transverse momentum,
thereafter. This shows that spectator and participant components
hardly interact after this time. However full equilibrium is not
attained even at 1000 AMeV. Interestingly, at all three incident
energies considered here, the anisotropy ratio $\langle R_{iso}
\rangle$ converge to same value. This indicates that equilibration
of participant matter in semicentral Ca+Ca collisions takes place
to nearly same extent, irrespective of the incident energy. These
findings are supported by the fact that in Plastic Ball
experiments, nearly the same baryonic entropy is produced
\cite{doss85} at different bombarding energies. The $R_{iso}$
value is not much altered unlike the mean directed transverse
momentum $\langle p^{dir}_{x} \rangle$. It may be mentioned that
momentum anisotropy ratio is still less than 1 with approximately
25 $\%$ anisotropy in momentum distribution. This also indicates
that surrounding spectator components has strong influence on the
evolution of participant matter. As a result, one can't have
always full stopping and thermalization even for the central
overlap region. As far as peripheral collisions are concerned, the
participant component, therefore, retains the \emph{transparency}
character in the longitudinal direction, and momentum distribution
of nucleons is far from that of thermalized source. This also
shows that nucleons are in quite highly non-equilibrium
environment that can not be described with single fireball model
\cite{gut}, or oversimplified hydrodynamical prescription
\cite{sim}.

In \ref{pt}, we display the momentum distribution of nucleons for
the single event of Ca+Ca collision obtained at 100 fm/c. The two
dimensional plane in the figure corresponds to longitudinal
momentum $p_{z}$ versus total transverse momentum
$p_{T}$$\left(=\sqrt{{p_{x}}^{2}+{p_{y}}^{2}}\right)$. Our model
predictions clearly show the impact of bombarding energy on
transverse expansion of the nuclear matter. This leads to increase
in $\langle p^{dir}_{x} \rangle$ with incident energy, as
mentioned earlier. One can observe appreciable \emph{transparency}
character of the nuclear matter in longitudinal direction. This
happens due to the fact total number of \emph{n-n} collisions
taking place in the medium are not able to keep the pace with
transverse expansion. Owing to this phenomenon, we have less
equilibrated matter with $\langle R_{iso} \rangle$ value slightly
less than unity.
\begin{figure}[!t]
\centering \vskip -0.60cm
\includegraphics [scale=0.5] {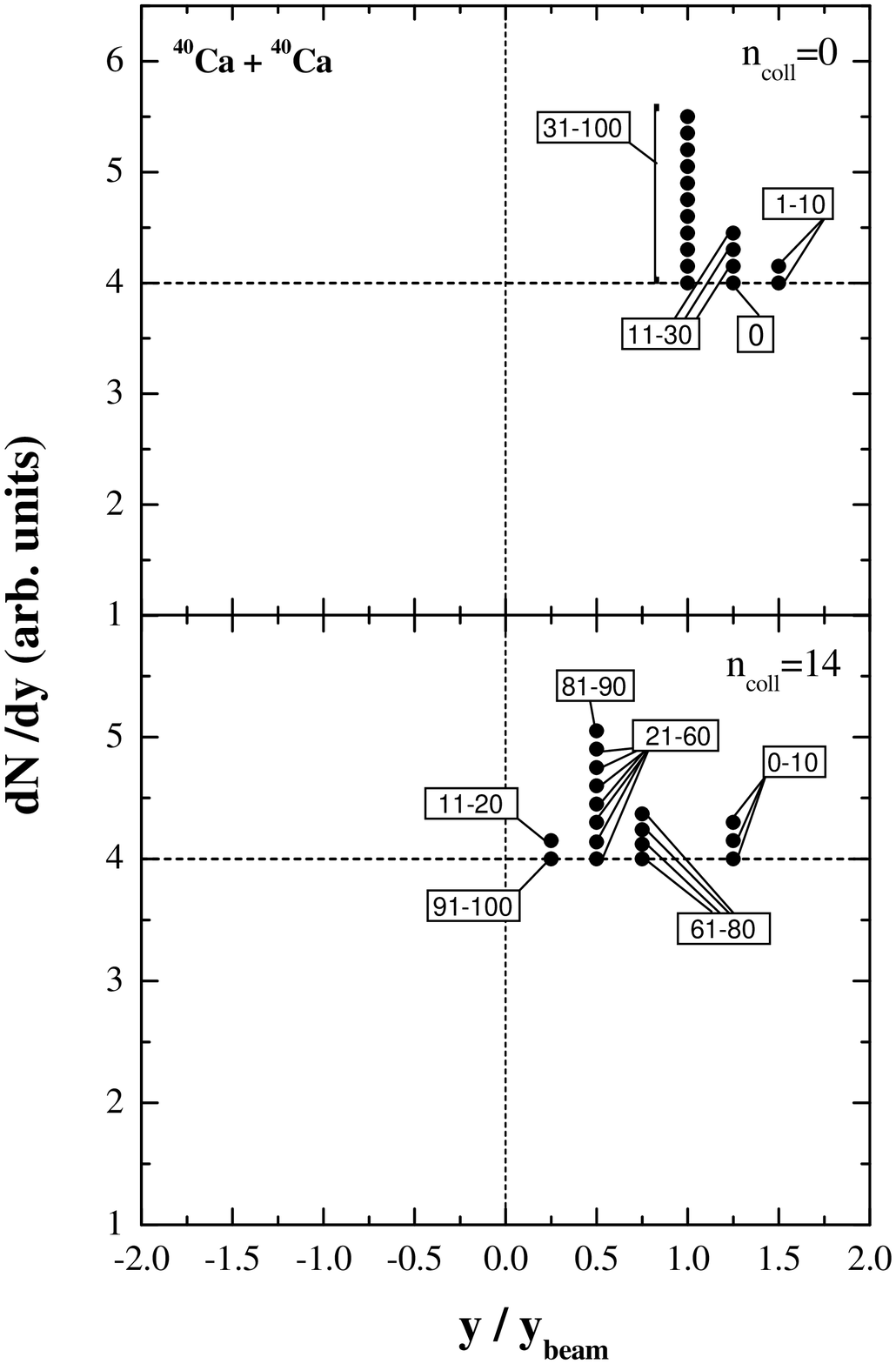} % Here is how to import EPS art
\vskip -0.5cm \caption{The rapidity distribution $dN/d$y vs
normalized rapidity y/y$_{beam}$ in the nucleus-nucleus c.m. frame
for the evolution of single spectator (top) and participant
(bottom) nucleons. Results are shown here for the single event of
Ca (400 AMeV)+ Ca collision with $b/b_{max}$=0.3.} \label{rap}
%\vskip -0.4 cm
\end{figure}
Next, we try to understand the stopping pattern of the nucleons
facing maximal and minimal number of collisions by analyzing the
variation in their rapidities at different time steps. Figure
\ref{rap} shows the variation of normalized rapidity y/y$_{beam}$
of single spectator nucleon, facing zero collision ($n_{coll}=0$)
and participant nucleon ($n_{coll}=14$) followed for the time span
of 100 fm/c. We have displaced here the symbols corresponding to
same rapidity bin but obtained at different time steps along the
y-axis, though all these points have same probability of
occurrence. This is done to make the variation in the longitudinal
rapidity with time more vivid. The numbers in the boxes imply the
time interval for which the nucleon stays in particular rapidity
bin. At the start of the reaction, both particles are seen at
projectile beam rapidity y$_{beam}$. With advent of the collision
process, participant nucleons (suffering 14 collisions) are
stopped around \emph{mid-rapidity} with y/y$_{beam}$=0.25. On the
other side, the spectator nucleons (as shown in the upper panel)
still stay in the higher rapidity regime with y/y$_{beam}$=1.
Again, one can also notice that participant nucleon suffers more
fluctuations in its longitudinal rapidity. The spectator particle,
however, remains in the projectile's rapidity regime for most of
the time.
\begin{figure}[!t]
\centering \vskip -1.14cm
\includegraphics [scale=0.5, trim=20 10 0 0] {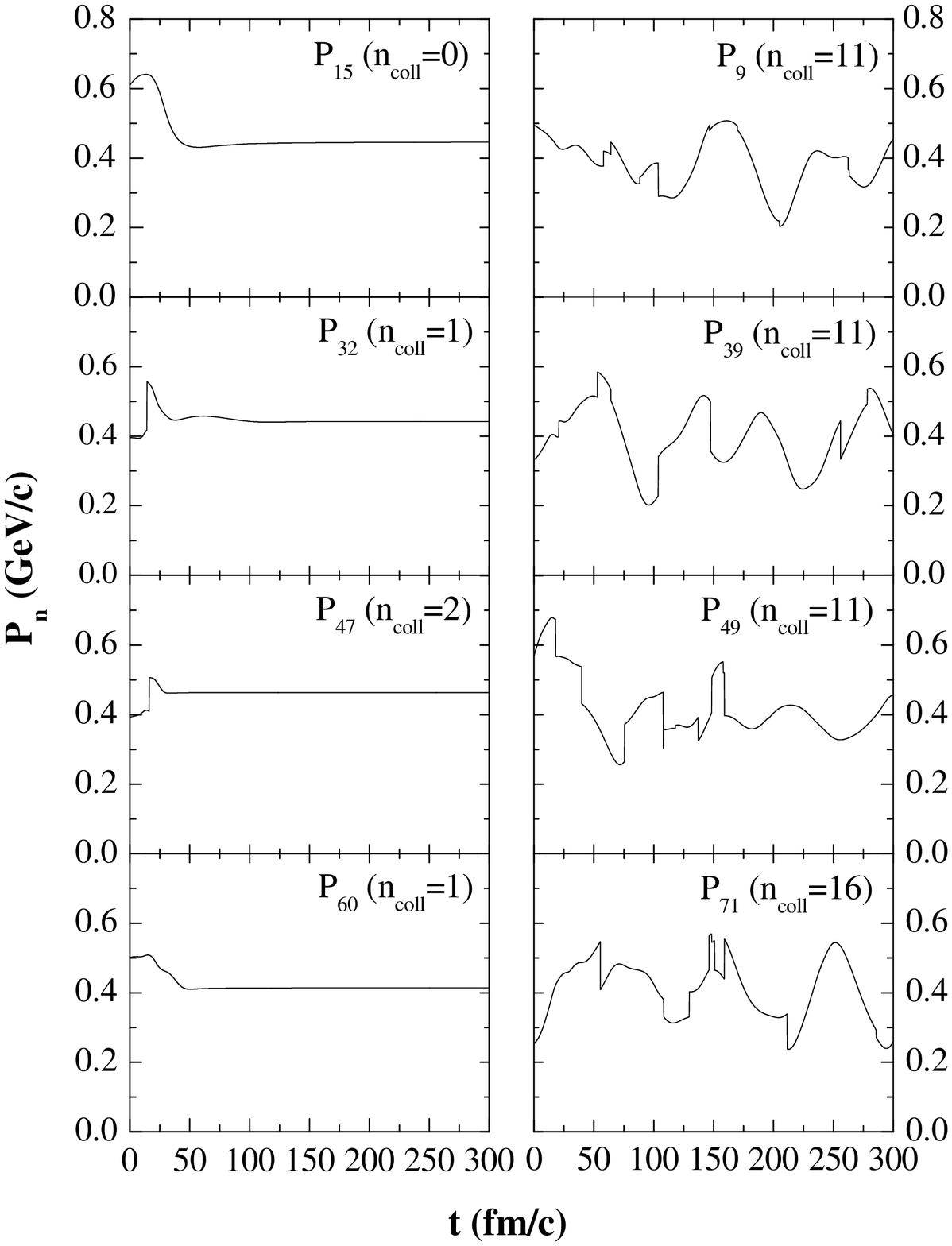} % Here is how to import EPS art
\vskip -0.7cm \caption{The time evolution of total momentum
$P_{n}$ (in GeV/c) of randomly chosen nucleons facing minimal
collisions (left) and maximal collisions (right).}
%\vskip -0.4 cm
\label{pn}
\end{figure}
The temperature reached in HI reactions at intermediate energies
can be as high as 70-80 MeV \cite{rkp94}. Nucleons feeling such
hot environment would behave differently than those found in
relatively cold spectator zones. It is interesting to follow the
space-time characteristics of these nucleons facing maximal
(\emph{i.e.} participants) and minimal (\emph{i.e.} spectators)
number of collisions. This study can be of importance to
understand characteristics of hot participant matter and hadronic
interactions as well. For this study, we randomly choose the
nucleons from target and projectile in $^{40}Ca$ (400 AMeV)+
$^{40}Ca$ collision. The left panel of \ref{pn}, shows the time
evolution of total momentum $P_{n}$ of individual spectator
nucleons ($P_{15},~P_{32},~P_{47},~P_{60}$). Similarly, the
evolution of participant nucleons
($P_{9},~P_{39},~P_{49},~P_{71}$) facing maximal collisions is
shown in the right panel.

\begin{figure*} [!t]
\centering \vskip -5.6cm
\includegraphics* [scale=0.65, trim=5 10 0 0] {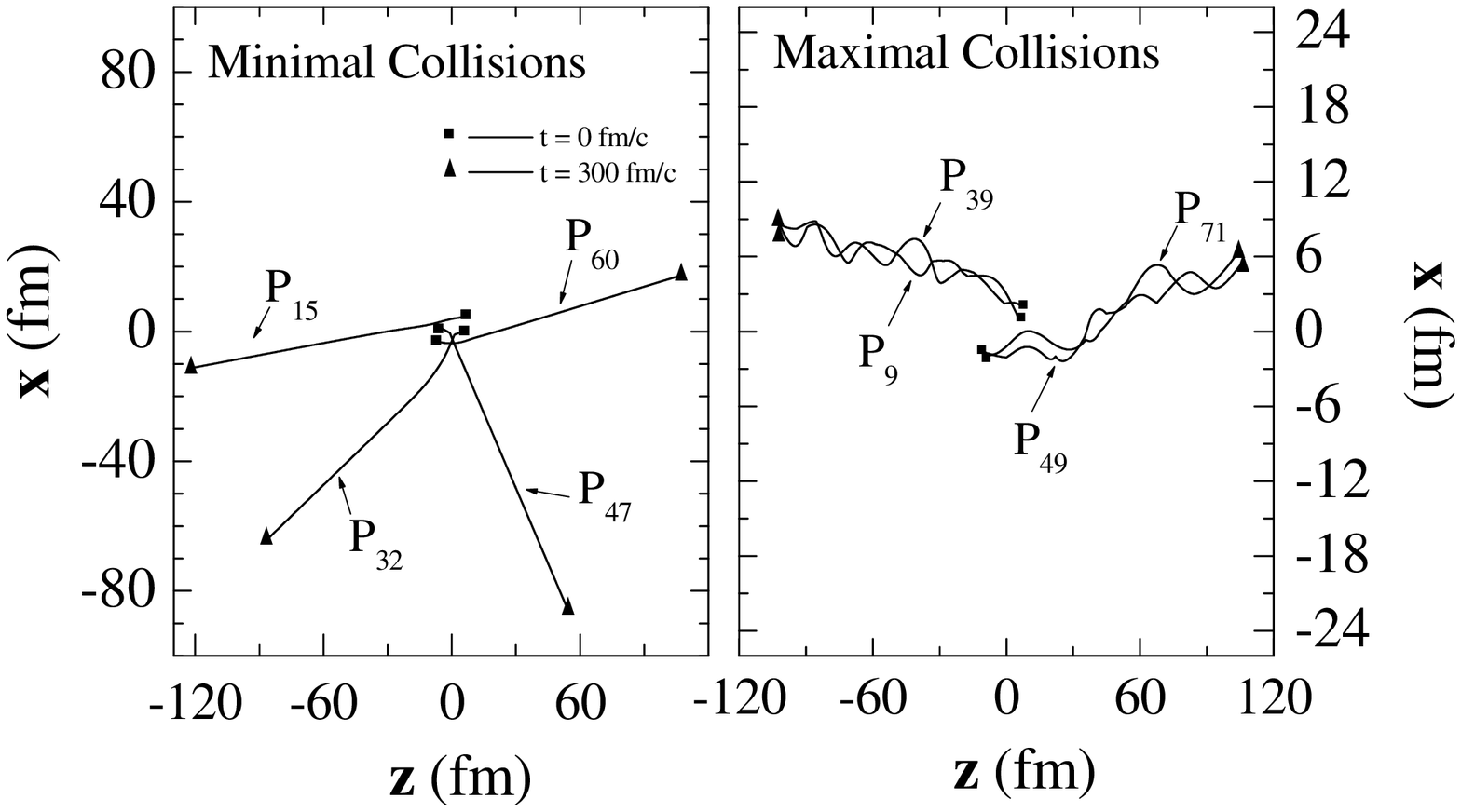} % Here is how to import EPS art
\vskip -4.7cm \caption{The trajectories of the randomly chosen
nucleons facing minimal (left panel) and maximal (right)
collisions followed for the time span of 300 fm/c. The
calculations are shown in the $x$-$z$ plane for the single event
of Ca (400 AMeV)+ Ca collision, with $b/b_{max}$=0.3.}
\label{traj}
%\vskip -0.4 cm
\end{figure*}

The particles behaving like spectators aren't expected to feel the
hot environment even during the violent stage of the reaction.
Owing to this, one obtains minimum fluctuations in the particle's
momentum during the collision process. On the other hand,
participant nucleons facing maximal number of collisions, feel the
hot environment as indicated by continuously change in their
momenta during the collision. The participant nucleons, therefore,
suffer more number of fluctuations in momentum space. We also aim
to analyze the behavior of these nucleons in coordinate space.
Figure \ref{traj} displays the trajectories of these nucleons in
$x$-$z$ plane followed for the time span of 300 fm/c starting from
the initial contact between the projectile and target nuclei.
Since nucleons facing minimal number of collisions are knocked-out
quite earlier, they traverse nearly straight line paths in ${\cal
R}_3$-space. Contrary to this, the participant nucleons (as shown
in the right panel) face large fluctuations along their
trajectories as expected. These particles remain in the hot zone
for appreciably long time unlike spectator nucleons which are
knocked-out of the reaction zone during the early stage of the
reaction.

\section{\label{summary}Summary}
Summarizing, we have presented an exclusive analysis of the
semicentral Ca+Ca reactions at incident energies 400, 600 and 1000
AMeV within quantum molecular dynamics approach. Our findings
showed that incident energy of the projectile strongly influences
the observables \emph{viz.} average nucleonic density, allowed
collision rate as well as the transverse expansion of nuclear
matter. Further, the transparency effect in the nuclear medium
also seems to get enhanced with the incident energy.
Interestingly, the degree of equilibrium achieved by the
participant zone remains insensitive to the range of projectile
energies considered in this work. The full equilibration is not
always there even for the participant zone. A detailed analysis of
trajectories traversed by the nucleons depicted that spectator and
participant nucleons behave differently in the coordinate and
momentum space. The nucleons facing maximal collisions \emph{i.e.}
participant nucleons suffered extensive fluctuations in their
momentum and spatial coordinates. The spectator nucleons, on other
hand, depart from the hot reaction zone quite earlier and traverse
nearly the straight line trajectories. \\
%%%%%%%%%%%%%%%%%%%%%%%%%%%%%%%%%%%%%%%%%

\end{document}